%% file: main.tex
\documentclass[sigchi]{acmart}

\AtBeginDocument{%
  \providecommand\BibTeX{{%
    \normalfont B\kern-0.5em{\scshape i\kern-0.25em b}\kern-0.8em\TeX}}}

\usepackage[linesnumbered, ruled, vlined]{algorithm2e}
\usepackage{amsthm}
\usepackage{hyperref}
\usepackage{float}
\usepackage{amssymb}
\usepackage{multirow}
\usepackage{stfloats}

\copyrightyear{2020}
\acmYear{2020}
\setcopyright{acmlicensed}\acmConference[WebSci '20]{12th ACM Conference on Web Science}{July 6--10, 2020}{Southampton, United Kingdom}
\acmBooktitle{12th ACM Conference on Web Science (WebSci '20), July 6--10, 2020, Southampton, United Kingdom}
\acmPrice{15.00}
\acmDOI{10.1145/3394231.3397903}
\acmISBN{978-1-4503-7989-2/20/07}

\begin{document}

\title{HPRA: Hyperedge Prediction using Resource Allocation}

\author{Tarun Kumar}
\authornote{Both authors contributed equally to this research.}
\email{tkumar@cse.iitm.ac.in}
\orcid{1234-5678-9012}
\affiliation{
  \institution{Robert Bosch Center for Data Science and AI\\Department of Computer Science and Engineering\\ IIT Madras}
}

\author{K Darwin}
\authornotemark[1]
\email{darwinreddy.k@gmail.com}
\affiliation{
  \institution{Robert Bosch Center for Data Science and AI\\Department of Computer Science and Engineering\\ IIT Madras}
}

\author{Srinivasan Parthasarathy}
\email{srini@cse.ohio-state.edu}
\affiliation{Department of Computer Science and Engineering\\ The Ohio State University}

\author{Balaraman Ravindran}
\email{ravi@cse.iitm.ac.in}
\affiliation{
  \institution{Robert Bosch Center for Data Science and AI\\Department of Computer Science and Engineering\\ IIT Madras}
}

\renewcommand{\shortauthors}{Tarun and Darwin, et al.}

\input{sections/abstract.tex}


\keywords{hypergraph, hyperedge prediction, resource allocation}

\maketitle

\input{sections/introduction.tex}
\input{sections/preliminaries.tex}
\input{sections/method.tex}
\input{sections/evaluation.tex}
\input{sections/conclusion.tex}

\bibliographystyle{ACM-Reference-Format}
\newpage
\bibliography{references}
\end{document}

%% file: sections/abstract.tex
\begin{abstract}
Many real-world systems involve higher-order interactions and thus demand complex models such as hypergraphs. For instance, a research article could have multiple collaborating authors, and therefore the co-authorship network is best represented as a hypergraph. In this work, we focus on the problem of hyperedge prediction. This problem has immense applications in multiple domains, such as predicting new collaborations in social networks, discovering new chemical reactions in metabolic networks, etc. Despite having significant importance, the problem of hyperedge prediction hasn't received adequate attention, mainly because of its inherent complexity. In a graph with $n$ nodes the number of potential edges is $\mathcal{O}(n^{2})$, whereas in a hypergraph, the number of potential hyperedges is $\mathcal{O}(2^{n})$. To avoid searching through the huge space of hyperedges, current methods restrain the original problem in the following two ways. One class of algorithms assume the hypergraphs to be $k$-uniform where each hyperedge can have exactly $k$ nodes. However, many real-world systems are not confined only to have interactions involving $k$ components. Thus, these algorithms are not suitable for many real-world applications. The second class of algorithms requires a candidate set of hyperedges from which the potential hyperedges are chosen. In the absence of domain knowledge, the candidate set can have $\mathcal{O}(2^{n})$ possible hyperedges, which makes this problem intractable. More often than not, domain knowledge is not readily available, making these methods limited in applicability. We propose \textit{HPRA - Hyperedge Prediction using Resource Allocation}, the first of its kind algorithm, which overcomes these issues and predicts hyperedges of any cardinality without using any candidate hyperedge set. \textit{HPRA} is a similarity-based method working on the principles of the resource allocation process. In addition to recovering missing hyperedges, we demonstrate that \textit{HPRA} can predict future hyperedges in a wide range of hypergraphs. Our extensive set of experiments shows that HPRA achieves statistically significant improvements over state-of-the-art methods.
\end{abstract}

%% file: sections/introduction.tex
\section{INTRODUCTION}
Complex systems are often characterized by several components that interact with one another in multiple ways. Such systems are encountered in various domains such as brain networks in biology \cite{van2013network}, railway networks in transportation \cite{satchidanand2014studying}, multiple computers connected in the internet \cite{funel2018analysis}, humans involved in multifaceted relationships \cite{menichetti2014weighted}, etc. These systems are often modeled as graphs where nodes represent the components, and edges represent their interactions. Though graph-modeling is backed by rigorous graph theory, it assumes that the components of the system can only be involved in pairwise interactions. This assumption may lead to an over-simplification of systems with higher-order interactions, such as multiple web pages connected to a single web page \cite{berlt2007hypergraph}, multiple metabolites involved in a reaction \cite{shen2018genome}, a set of proteins forming a protein complex \cite{klamt2009hypergraphs}, several cities connected by a train \cite{satchidanand2014studying}, several people collaborating for a project \cite{han2009understanding}, etc. These systems can be more accurately modeled using hypergraphs where nodes represent the interacting components, and hyperedges capture higher-order interactions.

Recently, hypergraph modeling has been used to solve problems related to clustering \cite{li2017inhomogeneous}, inference \cite{yadati2018hypergcn, feng2018learning}, motif counting \cite{das2019fast}, centrality \cite{roy2015measuring, benson2019three}, page/image reputation in web \cite{berlt2007hypergraph, yu2014click}, etc.
In this paper, we focus on a less explored but extremely significant problem of hyperedge prediction. Recent studies have shown remarkable performance in edge (link) prediction \cite{liben2003link, wang07, zhang2018link, kim2018efficient} in graphs. Predicting future edges in temporal graphs is also an active area of research \cite{huang2009time,roopashree2014future,rahmaida2019predicting,Wang17}. Unlike edge prediction, hyperedge prediction has several bottlenecks, both semantically and computationally, making the problem more challenging.  The inherent complexity of hypergraphs hinders edge-prediction methods from directly predicting hyperedges. Unlike graphs, where an edge can connect only two nodes, a hyperedge can connect an arbitrary number of nodes. Thus, while in a graph, the maximum possible number of edges is $\mathcal{O}(n^{2})$, in a hypergraph, the maximum possible number of hyperedges is $\mathcal{O}(2^{n})$. Searching through this enormous space for potential hyperedges exacerbates the modeling and search challenge as compared to the traditional edge or link prediction. Existing hyperedge prediction methods typically sidestep this problem in the following ways:
\begin{itemize}
    \item \textbf{Restricting the hyperedge cardinality:} These methods are limited to work for $k$-uniform hypergraphs where each hyperedge connects exactly $k$ nodes \cite{music2010uniform}. However, most of the real-world hypergraphs have arbitrary-sized hyperedges, which makes these methods unsuitable for several practical applications.
    \item \textbf{Using a candidate hyperedge set:} These methods model the hyperedge prediction problem as a classification task \cite{zhang2018beyond, xu2013hyperlink}. These methods classify each hyperedge from the candidate set into a positive or negative class. However, these methods need intensive domain knowledge to construct the appropriate candidate hyperedge set, limiting their applicability in real-world systems. Moreover, the evaluation of such methods depends significantly on the choice of the candidate hyperedge set \cite{patil2020negative}.
\end{itemize}
Apart from the hyperedge prediction methods mentioned above, there exists a parallel line of work where the future group interactions are predicted by viewing them as a sequence of sets \cite{benson2016modeling,benson2018sequences}. These techniques work on the assumption that every future set (group interaction) is a resultant of exactly one set at previous timestamps. This assumption does not hold while predicting hyperedges in various real-world networks, making our problem more challenging.

We propose a \textit{resource allocation} based method \cite{lu2011linkRA, zhou2009predicting}, \textit{HPRA: Hyperedge Prediction using Resource Allocation}\footnote{Code is available at - https://github.com/darwk/HyperedgePrediction}, which predicts novel hyperedges of any cardinality. The major contributions in this paper are:
\begin{enumerate}
    \item We propose a computationally-efficient hyperedge prediction model, \textit{HPRA}, which can predict novel hyperedges without using any candidate set.
    \item We propose a variant of \textit{HPRA}, which can be used in conjunction with a candidate hyperedge set. 
    \item We show that \textit{HPRA} can predict future hyperedges in addition to recovering missing hyperedges in a wide range of hypergraphs.  
    \item Our comprehensive set of experiments demonstrates that \textit{HPRA} significantly outperforms the state-of-the-art methods in terms of widely used metrics such as \textit{area under the precision-recall curve (AUC)} and \textit{precision}.
\end{enumerate}

%% file: sections/preliminaries.tex
\begin{figure*}[ht!]
    \centering
    \includegraphics[scale=0.5, width=15cm]{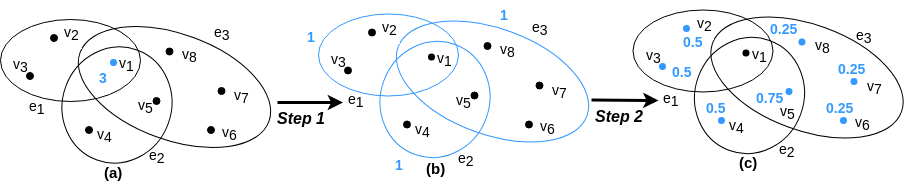}
    \caption{Example illustrating resource transfer directly between nodes in a hypergraph.}
    \label{HRA_Example}
\end{figure*}

\section{PRELIMINARIES}
Here, we introduce the mathematical notations used in the rest of the paper, followed by link prediction algorithms, specifically, resource allocation. 
\subsection{Hypergraphs}
A hypergraph is represented by a tuple $G = (V, E, w)$; where $V$ is the set of $n$ nodes (or vertices) and $E$ is the set of $m$ hyperedges. Each hyperedge $e$ has a positive weight $w(e)$ associated with it.
While in a traditional graph, an edge connects two nodes, a hyperedge can connect an arbitrary number of nodes. Degree of node $v$ is defined as $d(v) =  \sum_{e \in E,v \in e}w(e)$ and $N(v)$ is a set containing the one-hop neighbors of node $v$ (nodes of hyperedges, $v$ is part of). 
For a hyperedge $e$, its degree is defined as $\delta(e) = |e|$.  Incidence matrix $H$ is a $n \times m$ matrix with entries $h(v,e) = 1$ if $v \in e$, and $0$ otherwise. $D_v \in \mathbb{R}^{n \times n}$, $D_e \in \mathbb{R}^{m \times m}$ and $W \in \mathbb{R}^{m \times m}$ are the diagonal matrices containing node degrees, hyperedge degrees and hyperedge weights respectively. Then the adjacency matrix of hypergraph G is defined as\cite{hadleyengenvector}: $A = HWH^{T} - D_{v}$
\\
\textbf{Node Degree Preserving Reduction \cite{kumar2018hypergraph}}: During clique reduction, the node degree of a vertex $v$ is over counted by a factor of $(\delta(e)-1)$ for each hyperedge containing $v$. To preserve the node degree, we can scale down each $w(e)$ by a factor of $(\delta(e)-1)$. This results in the following adjacency matrix,
\begin{equation*}
    A_{ndp} = HW(D_{e} - I)^{-1}H^{T} - D_{v}
\end{equation*}

In our discussions, we assume $W$ to be an \textit{identity matrix} but our proposed approach is applicable to any $W$ with positive weights.

\subsection{Link Prediction using Similarity-Based Algorithms}

Several kind of real-world networks such as social networks, web networks are known to exhibit the property of \textit{homophily}, which states that similar nodes are more likely to connect in future than dissimilar nodes \cite{mcpherson2001birds, sarkar2011theoretical}. In accordance with this, similarity-based algorithms are broadly used for edge prediction in graphs. In a typical similarity-based algorithm, a similarity score is defined for node-pairs of a graph. Based on the similarity score, all the possible edges are ranked, and the top-ranked edges are chosen as the potential edges. Despite being a simple framework, defining an appropriate similarity score to capture the node similarities is a nontrivial task. One group of existing works use the node attributes to define the similarity score for node-pairs \cite{LinCommonAttributes}, but are restricted to attributed graphs. Without such restriction, another group of methods define similarity scores solely based on the network structure and are termed as structural similarity scores. Popular structural similarity scores are \textit{common neighbors} \cite{newman2001clustering}, \textit{resource allocation} \cite{zhou2009predicting} and \textit{katz index} \cite{Katz1953}. We elaborate on these scores in the later sections.

\subsection{Resource Allocation (RA)}
Motivated by the resource allocation process in networks \cite{georgiadis2006resource}, \textit{RA} score \cite{zhou2009predicting} for a node-pair $(x,y)$,  which are \textit{not directly connected} is defined as:
\begin{equation*}
            RA_{x y} = \sum_{z \in N(x) \cap N(y)}\frac{1}{d(z)}
\end{equation*}
To illustrate resource allocation in a simple way, assume node $x$ has a resource amount of $d(x)$ units allocated to it. Node $x$ transfers its resource to node $y$ through common neighbors, who act as transmitters in the following way; Node $x$ uniformly distributes its resource to all its neighbors, resulting in each neighbor of $x$ getting a unit of resource. Following this, neighbors of $x$ uniformly transfer their unit resource to their neighbors. The resource that node $y$ receives from node $x$ is defined as the resource allocation score for pair $(x,y)$. A higher amount of resource transferred between two nodes signifies a higher similarity among those nodes.

%% file: sections/method.tex
\section{METHOD}
Similar to the traditional link prediction methodology\cite{wang07}, a hyperedge can be predicted by exploring all possible hyperedges in the search space and ranking them. However, such an approach is infeasible as the hypergraph search space is vast $\mathcal{O}(2^{n})$ - unlike graphs, where the search space is $\mathcal{O}(n^{2})$. 
We overcome the difficulty of vast search space by pruning the space in steps. In this section, we first define the \textit{Hypergraph Resource Allocation (HRA)} index. Using \textit{HRA}, we define \textit{Node-Hyperedge Attachment Score (NHAS)}, which captures the likelihood of a node being a member of the new hyperedge. We then present \textit{HPRA: Hyperedge Prediction using Resource Allocation} to predict novel hyperedges.

\begin{figure*}[ht!]
    \centering
    \includegraphics[height=7cm, width=14cm]{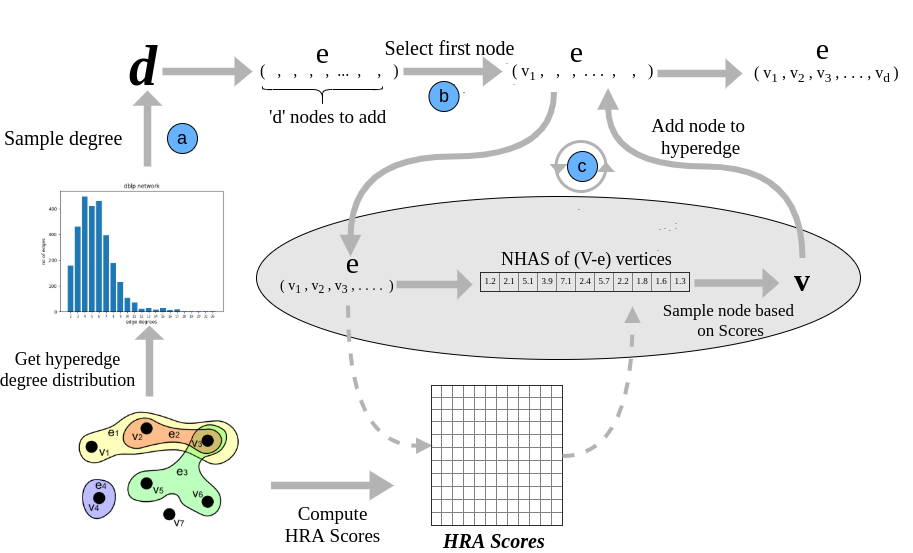}
    \caption{An Illustration of HPRA.  (a) cardinality of new hyperedge \textit{d} is sampled proportional to hyperedge degree distribution. (b) first node is chosen based on Preferential Attachment and added to \textit{e}. (c) subsequent nodes are sampled from set V$\setminus$e based on the \textit{NHAS} scores and added to \textit{e}. This step is repeated until \textit{d} nodes are added to \textit{e}}
    \label{Model}
\end{figure*}

\subsection{Hypergraph Resource Allocation (HRA)}

\textit{HRA} index is motivated by the \textit{RA} index defined for graphs. \textit{RA} index is defined for the node-pairs $(x,y)$ which are not directly connected, as these node-pairs are potential new edges. Unlike graphs, new hyperedges can have nodes which are already connected in the hypergraph.  
Thus, node $x$ can transfer its resources to $y$ by either a direct connection $(HRA_{direct})$ or via common neighbors $(HRA_{indirect})$. To determine $HRA_{direct}$, assume node $x$ has a resource of $d(x)$ units. Node $x$ uniformly distributes its resources to all hyperedges that include $x$. In the next step, the resource allocated to each hyperedge is uniformly distributed to the nodes of rhe hyperedge apart from $x$. The amount of resource node $y$ receives directly from node $x$ is given by $HRA_{direct}(x, y) = \sum_{e, s.t., x, y \in e}\frac{1}{\delta (e) - 1}$ (this is equal to $A_{ndp}(x,y)$ \cite{kumar2018hypergraph}). Figure \ref{HRA_Example} illustrates the direct transfer of resource between nodes in a toy hypergraph. Initially, $v_1$ distributes its resources to all three hyperedges uniformly, as shown in Figure \ref{HRA_Example}(b). In the next step, the unit resource allocated to each hyperedge is uniformly distributed to its nodes excluding $v_1$.

To determine $HRA_{indirect}$, assume node $z$ is a common neighbor of $x$ and $y$. Node $z$ receives $HRA_{direct}(x, z)$ amount of resource from $x$ and then distributes it to all its neighbors. Node $y$ being a neighbor of $z$, receives $HRA_{direct}(x,z)\times \frac{1}{d(z)}\times HRA_{direct}(z,y)$ amount of resource. Total resource received by $y$ through all common neighbors is given by:
{\scriptsize{
\begin{equation*}\label{hra-common-neighbor}
HRA_{indirect}(x,y) = \sum_{\tiny{z \in N(x) \cap N(y)}}{HRA_{direct}(x,z)\times \frac{1}{d(z)}\times HRA_{direct}(z,y)}
\end{equation*}}
}
Combining $HRA_{direct}$ score with $HRA_{indirect}$, we define the similarity between $x$ and $y$ as,

\begin{equation*}
            HRA_{x y} = HRA_{direct}(x,y) + HRA_{indirect}(x,y)
\end{equation*}

Notice that the \textit{HRA} computation depends only on the local neighborhood. Thus, $HRA$ can be computed efficiently for very large networks. 

\subsection{Node-Hyperedge Attachment Score (NHAS)}
Social networks are known to possess \textit{homophily}; for instance, consider a musical band $e$ looking for a $guitarist$. Assume guitarists $x$ and $y$ are known to be equally good, and guitarist $x$ has previously worked with few members from $e$. Then guitarist $x$ is more likely to be part of $e$ as compared to $y$. Here, the musical band represents a hyperedge and band members are the nodes. Following the principles of \textit{homophily} and using \textit{HRA} to capture node-node similarity, we formally define $NHAS$ as follows:
 \begin{equation*}
            NHAS_{x,e} = \frac{1}{|e|}\Big(\sum_{y \in e}HRA_{x y}\Big)
\end{equation*}

\subsection{HPRA: Hyperedge Prediction using Resource Allocation}

The problem of hyperedge prediction can be disintegrated into the following sub-problems:
\begin{enumerate}
    \item What should be the cardinality of the new hyperedge?
    \item Once the hyperedge-cardinality is determined, which node should be the first member of the new hyperedge?
    \item What are the other nodes that should be part of the new hyperedge?
\end{enumerate}
We propose \textit{HPRA} (Algorithm \ref{alg:HPRA}), which addresses the aforementioned problems in the following ways, and also depict the same in Figure \ref{Model}:
\begin{enumerate}
    \item A hyperedge prediction algorithm is expected to preserve the structural properties of the hypergraph \cite{guo2016evolution}. One such structural property is hyperedge degree distribution. To preserve it, the cardinality of the new hyperedge has to be \textit{in line} with the observed hyperedge degree distribution. Therefore, we sample the cardinality of the new hyperedge from the observed hyperedge degree distribution. In other words, the higher the number of observed hyperedges with degree $d$, the higher the probability that the new hyperedge cardinality is $d$.
    
    However, there is a possibility of not encountering the hyperedges of a specific cardinality in the hypergraph. To predict hyperedges with such cardinalities, we smoothen the hyperedge degree distribution by following a Laplace smoothing operator and work with the resultant distribution \cite{valcarce2016additive}. Thus, we always have a \textit{fail-safe} probability to handle missing cardinalities in the hypergraph.
    \item More often than not, social \cite{guo2016non} and web networks \cite{kunegis2013preferential} evolve by following the principles of preferential attachment, i.e., nodes with a higher degree are more likely to form new links. Following this, once the cardinality $d$ of new hyperedge is determined, we choose the first member of the hyperedge with probability proportional to the node degrees.
    \item As the new hyperedge $e$ is initialized with one node $v_{new}$, we compute \textit{NHAS} (Algorithm \ref{alg:NHAS}) of all the remaining nodes and the new hyperedge $e$. We sample a node from the set $V\setminus e$ proportional to \textit{NHAS} and add it to $e$. We repeat this step, until $\delta(e)$ becomes equal to $d$. 
\end{enumerate}

\begin{algorithm}[!ht]
\caption{Hyperedge Prediction using Resource Allocation (HPRA)}
\label{alg:HPRA}
\SetAlgoLined
\DontPrintSemicolon

\KwIn{\ Hypergraph Incidence Matrix \textit{H}, Node set \textit{V}, Hyperedge Degree Distribution \textit{HDD}}
\KwOut{Predicted Hyperedge \textit{e}}
// Initialize hyperedge \textit{e} \;
$e \gets \{\}$

// Sample hyperedge degree from \textit{HDD} \;
$d \gets get\_degree(hyperedge\_degrees, prob=HDD)$

// Select first node using Preferential Attachment \;
$v_{new} =  get\_node(V, prob=node\_degrees)$

$e.add(v_{new})$

\While {$size(e) < d$}{
    // Compute \textit{NHAS} for remaining nodes\;
    $scores \gets NHAS(e, V)$
    
    // Select a node based on \textit{NHAS} \;
    $v \gets get\_node(V, prob=scores)$
    
    $e.add(v)$
}
\end{algorithm}

\begin{algorithm}[!ht]
\caption{Node-Hyperedge Attachment Scores}
\label{alg:NHAS}
\SetAlgoLined
\DontPrintSemicolon

\KwIn{\ Edge \textit{e}, Node set \textit{V}, HRA score matrix \textit{HRA}}

\KwOut{Node-Hyperedge Attachment Scores \textit{scores})}

// Initialize scores \;
$scores \gets zeroes(size(V))$

// Compute NHAS for each node in \textit{V} $\setminus$ \textit{e} \;
\For {$v_i \ in \ V$}{ 
    \If {$v_i \ not \ in \ e$}{
        \For {$v_j \ in \ e$}{
            $scores[i] \gets scores[i] + HRA(v_i, v_j)$
        }
        $scores[i] \gets \frac{1}{size(e)}*(scores[i])$
    }
}
\end{algorithm}

%% file: sections/evaluation.tex
\begin{table*}[ht!]
\centering
\begin{tabular}{clcccc}
\hline
 &\textbf{Datasets} & \textbf{\# nodes} & \textbf{\# hyperedges} & \textbf{Avg hyperedge degree} & \textbf{Avg node degree} \\ 
\hline
(a) & Citeseer Co-reference & 1299 & 626 & 4.6102 & 2.2217 \\ 
(b) & Citeseer Co-citation & 1016 & 817 & 3.4198 & 2.7500 \\
(c) & Cora Co-reference & 1961 & 875 & 5.2594 & 2.3467 \\
(d) & Cora Co-citation & 1339 & 1503 & 3.0598 & 3.4579 \\ 
(e) & DBLP Co-authorship & 4695 & 2561 & 5.6177 & 3.0643\\
(f) & Movielens & 3893 & 4677 & 79.8751 & 95.9609 \\
(g) & HiggsTwitter & 9948 & 9605 & 47.7413 & 46.0951 \\
(h) & Amazon Co-view & 18565 & 10839 & 13.9063 & 8.1190 \\
(i) & Amazon Co-purchase & 24944 & 27675 & 41.7589 & 46.3309 \\
(j) & ArnetMiner Co-citation  & 21375 & 17300 & 4.1304 & 3.3429 \\
(k) & ArnetMiner Co-reference & 16620 & 26640 & 4.5388 & 7.2752 \\
\hline
\end{tabular}
\caption{Datasets Description}
\label{datasets}
\end{table*}

\section{Experiments}

We evaluate the performance of \textit{HPRA} on a broad range of networks. We propose new baselines by extending state-of-the-art similarity measures to be used with \textit{HPRA} framework by replacing the \textit{HRA} index. Before elaborating on the experimental setup, we first introduce the baselines and datasets used for evaluation.

\subsection{Baselines}

\textbf{Coordinated Matrix Minimisation (CMM) \cite{zhang2018beyond}}: CMM is based on matrix factorisation in adjacency space of hypergraph. It uses the EM algorithm to determine the presence or absence of candidate hyperedges.\\
\textbf{Spectral Hypergraph Clustering (SHC) \cite{zhou2007learning}}: SHC models the task of hyperedge prediction as a classification problem. Hypergraph Laplacian is used to classify the new hyperedges into positive or negative class.\\
\textbf{Common Neighbors (CN) \cite{newman2001clustering}, Katz \cite{Katz1953} }: CN and Katz are pairwise similarity indices for link prediction. CN is a local measure that assigns a similarity score based on the common neighbors of two nodes. Katz index is a global measure that captures the similarity between two nodes by considering paths connecting the nodes. A damping factor $\beta$ is used to assign higher importance to relatively shorter paths. $\beta$ is determined by searching over \{0.005,0.01,0.05,0.1,0.5\} using cross-validation.\\

\subsection{Datasets}

For our experiments, we only use the largest connected component of the network. Statistics of the datasets are shown in Table \ref{datasets}.
\\
\textbf{Cora, Citeseer \cite{sen2008collective} and ArnetMiner \cite{Tang:08KDD}}: We built two networks from each dataset; co-citation and co-reference where a node represents a paper. In a co-citation network, a hyperedge connects papers cited together. Similarly, in a co-reference network, if a set of papers refer to the same paper, they are connected by a hyperedge. 
\\
\textbf{HiggsTwitter \cite{de2013anatomy}}: This dataset captures messages posted on  Twitter about the Higgs boson discovery. We built a social network, where a node represents a person and hyperedge connects all people following the same person.
\\
\textbf{DBLP \cite{ley2002dblp}}: This is a co-authorship dataset. Here, a node represents an author and hyperedge connects authors of the paper.
\\
\textbf{Movielens \cite{movielens}}: This is a multi-relational dataset, where nodes represent movies and a hyperedge connects movies directed by the same individual.
\\
\textbf{Amazon Product Metadata \cite{he2016ups}}: 
We used metadata of products from the video games category and built two networks; co-view and co-purchase. In both networks, nodes represent products. In a co-view network, a hyperedge connects products viewed by customers at the time of purchase. Similarly, in a co-purchase network, a hyperedge connects products purchased together by customers.

\subsection{Evaluation of HPRA}
In general, we are not aware of the missing hyperedges. Therefore, for experimentation, we randomly divide the hyperedges into two sets: Training set ($E^T$) and Missing set ($E^M$). By treating $E^T$ as the observed hyperedges, we try to predict hyperedges of $E^M$. To remove any unwanted bias, we partition the hyperedges into $K$ subsets. Every time we select one subset as $E^M$ and the remaining $K - 1$ subsets jointly as $E^T$. We repeat the cross-validation process $K$ times, with each of the $K$ subsets being used exactly once as the $E^M$. This way, each hyperedge is used for testing exactly once. However, this approach has a limitation that after splitting the hyperedge set, few nodes may not be connected to any other node in the $E^T$. It is not practical to expect the method to predict hyperedges having such nodes. Therefore, we remove these hyperedges from $E^M$. Once we have the final $E^T$ and $E^M$, we generate $|E^M|$ number of new hyperedges by treating $E^T$ as observed hyperedges using \textit{HPRA} and call it as the predicted hyperedges set ($E^P$).  We evaluate the performance of our algorithm by computing the Average F1 score \cite{yang2013overlapping}.\\ 
\begin{itemize}
\item \textbf{Average F1 Score}: We use this measure to quantify the closeness of predicted hyperedges to the missing hyperedge set. Average F1 score is the average of the F1-score of the best matching missing hyperedge to each predicted hyperedge and the F1-score of the best-matching predicted hyperedge to each missing hyperedge:
\begin{equation*}
    \begin{split}
        Average \ F1 \ Score &= \frac{1}{2}\Big(\frac{1}{|E^M|}\sum_{e_i \in E^M} F1(e_i,\hat{e}_{g(i)}) + \\
        & \qquad \enspace \frac{1}{|E^P|}\sum_{\hat{e}_i \in E^P} F1(e_{g'(i)},\hat{e}_i)\Big)
    \end{split}
\end{equation*}

where $g$ and $g'$ are defined as follows:
\begin{align*}
    g(i) &= argmax_j(F1(e_i, \hat{e}_j))\\
    g'(i) &= argmax_j(F1(e_j, \hat{e}_i))
\end{align*}
\end{itemize}

To compare \textit{HPRA} with Katz and CN, we use the respective pairwise scores instead of the \textit{HRA} score in our framework. For datasets (a) to (e), we used $5$-fold cross validation. For rest of the datasets, we used $10$-fold cross validation.

\begin{table*}[!ht]
\centering
\begin{tabular}{|l|c|c|c|}
\hline
\textbf{} & \textbf{Katz} & \textbf{CN} & \textbf{HPRA} \\ 
\hline
\hline
(a) & 0.1346 $\pm$ 0.0366 & 0.1221 $\pm$ 0.0259 & \textbf{0.1449 $\pm$ 0.0127} \\
\hline
(b) & 0.2570 $\pm$ 0.0219 & 0.2568 $\pm$ 0.0170 & \textbf{0.2949 $\pm$ 0.2030} \\
\hline
(c) & 0.1199 $\pm$ 0.0125 & 0.1024 $\pm$ 0.0177 & \textbf{0.1303 $\pm$ 0.0225} \\
\hline
(d) & 0.3644 $\pm$ 0.0110 & 0.3389 $\pm$ 0.0058 & \textbf{0.3866 $\pm$ 0.0075} \\
\hline
(e) & 0.2480 $\pm$ 0.0051 & 0.2215 $\pm$ 0.0073 & \textbf{0.2855 $\pm$ 0.0077} \\
\hline
(f) & 0.1050 $\pm$ 0.0007 & 0.1049 $\pm$ 0.0008 &  \textbf{0.1215 $\pm$ 0.0007} \\
\hline
(g) & 0.1472 $\pm$ 0.0071 & 0.1529  $\pm$ 0.0046
& \textbf{0.1921  $\pm$ 0.0090} \\
\hline
(h) & 0.1290 $\pm$ 0.0034 & 0.1469 $\pm$ 0.0072 &  \textbf{0.2218 $\pm$ 0.0061} \\
\hline
(i) & 0.1405 $\pm$ 0.0025 & 0.1565 $\pm$ 0.0033 & \textbf{0.2234 $\pm$ 0.0048} \\
\hline
(j) & 0.2256 $\pm$ 0.0059 & 0.2225 $\pm$ 0.0056 & \textbf{0.2495 $\pm$ 0.0058} \\
\hline
(k) & 0.2676 $\pm$ 0.0034 & 0.2530 $\pm$ 0.0031 & \textbf{0.2895 $\pm$ 0.0027} \\
\hline
\end{tabular}
\vspace{10pt}
\caption{Average F1 Scores of HPRA and baselines. First column represents datasets described in Table \ref{datasets}.}
\label{table:Avg F1 score}
\end{table*}

\begin{table*}[!ht]
\centering
\begin{tabular}{|c|c|c|c|c|c|}
\hline
\textbf{\quad\quad} & \textbf{CMM} & \textbf{SHC} & \textbf{Katz} & \textbf{CN} & \textbf{HPRA} \\ 
\hline
\hline
(a) & 0.2966 $\pm$ 0.0340 & 0.5884 $\pm$ 0.0382 & 0.8396 $\pm$ 0.0806 & 0.8347 $\pm$ 0.0153 & \textbf{0.9007 $\pm$ 0.0067} \\
\hline
(b) &  0.3824 $\pm$ 0.0707 & 0.7512 $\pm$ 0.0246 & 0.8831 $\pm$ 0.0167 & 0.8460 $\pm$ 0.0135 & \textbf{0.8999 $\pm$ 0.0145} \\
\hline
(c) & 0.4074 $\pm$ 0.0407 & 0.5487 $\pm$ 0.0170 & 0.8290 $\pm$ 0.0272 & 0.7877 $\pm$ 0.0281 & \textbf{0.8508 $\pm$ 0.0222} \\
\hline
(d) & 0.3662 $\pm$ 0.0057 & 0.8007 $\pm$ 0.0204 & \textbf{0.9374 $\pm$ 0.0076} & 0.9073 $\pm$ 0.0089 & 0.9227 $\pm$ 0.0099 \\
\hline
(e)  & 0.0716 $\pm$ 0.0270 & 0.8076 $\pm$ 0.0295 & 0.9892 $\pm$ 0.0092 & 0.9812 $\pm$ 0.0095 & \textbf{0.9898 $\pm$ 0.0076} \\
\hline
(f)  & 0.0607 $\pm$ 0.0316 & 0.6578 $\pm$ 0.0091 & 0.5683 $\pm$ 0.0974 & 0.9689 $\pm$ 0.0010 & \textbf{0.9936 $\pm$ 0.0016} \\
\hline
(g) & - & 0.6057 $\pm$ 0.0271 & 0.4763 $\pm$ 0.0644 & 0.8051 $\pm$ 0.0124 & \textbf{0.9874 $\pm$ 0.0021} \\
\hline
(h)  & - & 0.5701 $\pm$ 0.0122 & 0.5810 $\pm$ 0.0349 & 0.9859 $\pm$ 0.0035 & \textbf{0.9897 $\pm$ 0.0032} \\
\hline
(i)  & - & 0.5941 $\pm$ 0.0106 & 0.3801 $\pm$ 0.1438 & 0.9842 $\pm$ 0.0017 & \textbf{0.9979 $\pm$ 0.0009} \\
\hline
(j)  & - & 0.6637 $\pm$ 0.0093 & \textbf{0.9293 $\pm$ 0.0072} & 0.9126 $\pm$ 0.0059 & 0.9237 $\pm$ 0.0063 \\
\hline
(k) & - & 0.6059 $\pm$ 0.0064  & 0.8070 $\pm$ 0.0416 & 0.9080 $\pm$ 0.0034 & \textbf{0.9421 $\pm$ 0.0026} \\
\hline
\end{tabular}
\vspace{10pt}
\caption{AUC results of HPRA and aforementioned baselines. First column represents datasets described in Table \ref{datasets}. The missing entries correspond to experiments that did not complete even after 24 hours of execution.}
\label{table:auc}
\end{table*}

\begin{table*}[ht!]
\centering
\begin{tabular}{|c|c|c|c|c|c|}
\hline
\textbf{\quad\quad} & \textbf{CMM} & \textbf{SHC} & \textbf{Katz} & \textbf{CN} & \textbf{HPRA} \\ 
\hline
\hline
(a) & 0.0403 $\pm$ 0.0248 & 0.1235 $\pm$ 0.0401 & 0.6901 $\pm$ 0.0906 & 0.6666 $\pm$ 0.0599 & \textbf{0.7803 $\pm$ 0.131} \\
\hline
(b) & 0.1502 $\pm$ 0.0507 & 0.3721 $\pm$ 0.0308 & 0.8014 $\pm$ 0.0159 & 0.7344 $\pm$ 0.0279 & \textbf{0.8299 $\pm$ 0.0326} \\
\hline
(c) & 0.0588 $\pm$ 0.0282 & 0.1192 $\pm$ 0.0146 & 0.6972 $\pm$ 0.0374 & 0.6582 $\pm$ 0.0539 & \textbf{0.7615 $\pm$ 0.0231} \\
\hline
(d) & 0.1096 $\pm$ 0.0391 & 0.4580 $\pm$ 0.0414 & \textbf{0.8649 $\pm$ 0.0157} & 0.8391 $\pm$ 0.0161 & 0.8588 $\pm$ 0.0164 \\
\hline
(e)  & 0.0072 $\pm$ 0.0063 & 0.4787 $\pm$ 0.0525 & \textbf{0.9625 $\pm$ 0.0185} & 0.9152 $\pm$ 0.0237 & 0.9523 $\pm$ 0.0099 \\
\hline
(f)  & 0.0216 $\pm$ 0.0075 & 0.4445 $\pm$ 0.0132 & 0.3319 $\pm$ 0.0765 & 0.8722 $\pm$ 0.0077 & \textbf{0.9555 $\pm$ 0.0019} \\
\hline
(g) & - & 0.3905 $\pm$ 0.0132 & 0.3099 $\pm$ 0.0499 & 0.5554 $\pm$ 0.0129 & \textbf{0.9220 $\pm$ 0.0053} \\
\hline
(h)  & - & 0.4529 $\pm$ 0.0137 & 0.5699 $\pm$ 0.0354 & 0.9561 $\pm$ 0.0046 & \textbf{0.9698 $\pm$ 0.0069} \\
\hline
(i)  & - & 0.4161 $\pm$ 0.0161  & 0.3515 $\pm$ 0.1364 & 0.9577 $\pm$ 0.0030 & \textbf{0.9914 $\pm$ 0.0014} \\
\hline
(j)  & - &  0.4265 $\pm$ 0.0202 & \textbf{0.8669 $\pm$ 0.0114} & 0.8172 $\pm$ 0.0104 & 0.8394 $\pm$ 0.0112 \\
\hline
(k) & - & 0.4782 $\pm$ 0.0067 & 0.7242 $\pm$ 0.0356 & 0.6475 $\pm$ 0.0135 & \textbf{0.8581 $\pm$ 0.0031} \\
\hline
\end{tabular}
\vspace{10pt}
\caption{Precision results of HPRA and aforementioned baselines. First column represents datasets described in Table \ref{datasets}. The missing entries correspond to experiments that did not complete even after 24 hours of execution.}
\label{table:precision}
\end{table*}

\subsection{HPRA with a candidate hyperedge set (HPRA-CHS)}

To compare the performance against the methods which use a candidate set, we propose a variant of \textit{HPRA}. In \textit{HPRA-CHS}, we select the top $|E^M|$ hyperedges based on \textit{HRA} score as predictions. For a candidate hyperedge, \textit{HRA} score is computed by taking the average of all pairwise ($\frac{m(m-1)}{2}$) \textit{HRA} indices.

Similar to the above setting, for evaluating \textit{HPRA-CHS}, we divide the hyperedges into a Training set ($E^T$) and Missing set ($E^M$). We build a candidate set consisting of the missing set $E^M$ and a set of distractor hyperedges. Distractor hyperedges are generated randomly based on the hyperedge degree distribution of the network. In our experiments, the distractor hyperedges set is ten times the size of missing hyperedges set. We generalize the Katz and CN pairwise indices using a method similar to \textit{HRA}. We evaluate our method using two standard metrics, AUC (Table \ref{table:auc}) and Precision (Table \ref{table:precision}), similar to \cite{lu2011linkRA, zhang2018beyond}. In Tables, \lq-\rq correspond to experiments that did not complete even after 24 hours of execution on a 64GB, Intel Xeon processor.

\begin{itemize}
    \item \textbf{AUC}: AUC score can be interpreted as the probability that a randomly chosen missing hyperedge is assigned a higher score than a randomly chosen distractor hyperedge.
    \item \textbf{Precision}: Given the rank of the hyperedges in the candidate set, precision is defined as the ratio of actual missing hyperedges to the number of predicted hyperedges. That is to say, if we choose the top $L$ ones ($L$ is the size of missing hyperedges set) as predicted hyperedges, among which $L_m$ hyperedges are in missing hyperedge set, then precision is equal to ($\frac{L_m}{L}$).
\end{itemize}

\begin{figure*}[ht!]
    \centering
    \includegraphics[height=5cm, width=14cm]{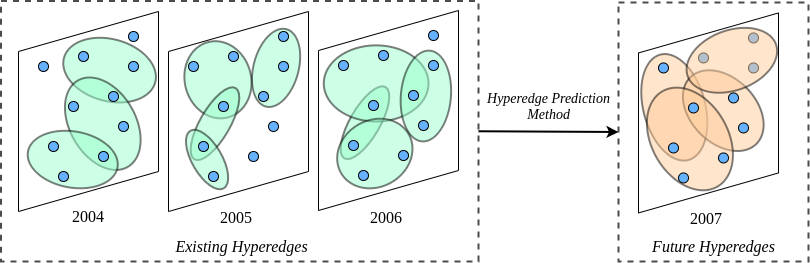}
    \caption{An illustration of the temporal dataset. Each graph represents the co-citation hypergraph in the respective year. Here, the task is to predict the hyperedges in the latter hypergraph (2007 in the figure) by using the information from previous years.}
    \label{fig:temporal-hypergraph}
\end{figure*}

\begin{table*}[ht!]
\centering
\begin{tabular}{cccccccc}
\hline
\multirow{2}{*}{} & \textbf{Existing hyperedges} & \textbf{Future hyperedges} & \# \textbf{nodes} & \# \textbf{existing} & \# \textbf{future} & \textbf{Avg hyperedge} & \textbf{Avg node} \\
& \textbf{(years)} & \textbf{(years)} & & \textbf{hyperedges} & \textbf{hyperedges} & \textbf{degree} & \textbf{degree} \\
\hline
(1) & 2000-2002 & 2003 & 10140 & 20234 & 6941 & 6.5527 & 13.0758 \\ 
(2) & 2001-2003 & 2004 & 11827 & 24018 & 7574 & 6.8275 & 13.8652 \\
(3) & 2002-2004 & 2005 & 13007 & 33452 & 15265 & 6.4968 & 16.7089 \\
(4) & 2003-2005 & 2006 & 16903 & 45090 & 17489 & 6.8317 & 18.2242 \\ (5) & 2004-2006 & 2007 & 22143 & 60265 & 20007 & 7.1386 & 19.4288 \\
\hline
\end{tabular}
\caption{Temporal ACM Cocitation Dataset Description.}
\label{temporal_datasets}
\end{table*}

\begin{table*}[ht!]
\centering
\begin{tabular}{|c||c|c|c||c|c|c|c|c||c|c|c|c|c|}
\hline
\multirow{2}{*}{\quad\quad} & \multicolumn{3}{c||}{\textbf{Average F1 score}} & \multicolumn{5}{c||}{\textbf{AUC}} & \multicolumn{5}{c|}{\textbf{Precision}} \\
\cline{2-14}
& \textbf{Katz} & \textbf{CN} & \textbf{HPRA} & \textbf{CMM} & \textbf{SHC} & \textbf{Katz} & \textbf{CN} & \textbf{HPRA} & \textbf{CMM} & \textbf{SHC} & \textbf{Katz} & \textbf{CN} & \textbf{HPRA}  \\ 
\hline
(1) & 0.3288 & 0.3016 & \textbf{0.3658} & 0.2508 & 0.5678 & 0.9330 & 0.9359 & \textbf{0.9531} & 0.3086 & 0.6124 & 0.8721 & 0.8797 & \textbf{0.9059} \\
\hline
(2) & 0.3174 & 0.2949 & \textbf{0.3531} & 0.2413 & 0.5592 & 0.9423 & 0.9442 & \textbf{0.9563} & 0.3132 & 0.6368 & 0.8745 & 0.8959 & \textbf{0.9167} \\
\hline
(3) & 0.3478 & 0.3192 & \textbf{0.3863} & - & 0.5378 & 0.9442 & 0.9449 & \textbf{0.9529} & - & 0.6211 & 0.8823 & 0.9090 & \textbf{0.9286} \\
\hline
(4) & 0.3305 & 0.3063 & \textbf{0.3729} & - & 0.5763 & 0.9543 & 0.9578 & \textbf{0.9720} & - & 0.6317 & 0.8930 & 0.9226 & \textbf{0.9393} \\
\hline
(5) & 0.3171 & 0.2927 & \textbf{0.3569} & - & 0.5814 & 0.9608 & 0.9650 & \textbf{0.9782} & - & 0.6415 & 0.8995 & 0.9284 & \textbf{0.9391} \\
\hline
\end{tabular}
\caption{AUC, Precision and Average F1 score results. First column represents datasets described in Table \ref{temporal_datasets}. The missing entries correspond to experiments that did not complete even after 24 hours of execution.}
\label{table:temporal_auc}
\end{table*}

\pagebreak

\subsection{Future hyperedge prediction with HPRA}
In this setting, we consider the task of predicting hyperedges of later years by using the previous years' hyperedges, as shown in Figure \ref{fig:temporal-hypergraph}. Here, the future hyperedge set consists of hyperedges only from a particular year, and the similarity scores are calculated using network snapshots from previous years. We use ACM co-citation dataset \cite{Tang:08KDD} for our experiments, and the dataset statistics are shown in Table \ref{temporal_datasets}. We evaluate both variants of HPRA- with and without a candidate hyperedge set, and report the results in Table \ref{table:temporal_auc}. 

\subsection{Results and Discussions}

From Table \ref{table:Avg F1 score}, \ref{table:auc}, \ref{table:precision} and \ref{table:temporal_auc}, following observations are evident:
\begin{itemize}
    \item \textit{HPRA} outperforms other baselines on most of the datasets and achieves highly competitive performance with the best results on the rest. In order to validate the results, we performed the \textit{paired t-tests} and observed the following 
    \begin{itemize}
        \item[-] In Table \ref{table:Avg F1 score}, HPRA-average F1 score is higher than the other methods with $p<0.05$ on all but  \textit{Citeseer Co-reference} and \textit{Cora Co-reference} datasets.
        \item[-] In Table \ref{table:auc}, Whenever HPRA performs the best, HPRA-average AUC score is higher than the other methods with $p<0.05$ on all the datasets except DBLP.
        \item[-] In Table \ref{table:precision}, Whenever HPRA performs the best, HPRA-average Precision score is higher than the other methods with $p<0.05$ on all the datasets.
    \end{itemize}
    \item None of the other baselines performed consistently well on all datasets, while HPRA is either the best performing or close to the best on all the datasets. 
    \item Though \textit{katz} has a reasonably good performance on many datasets, it fails to perform when hypergraphs have high average hyperedge and node degrees. One possible reason could be, in such hypergraphs, even a small damping factor may involve a large proportion of the hypergraph in score calculation, which may lead to identical similarity scores for multiple node-pairs.
    \item \textit{HPRA} performs remarkably well on \textit{HiggsTwitter} dataset, while other methods perform poorly. One distinguishing characteristic of this dataset is that the nodes with a low degree are part of hyperedges with high cardinality, and the nodes with a high degree participate in low cardinality hyperedges. This distinct pattern of nodes' participation causes the CN approach to perform poorly. High average node degree and hyperedge cardinality of the hypergraph introduces unwanted influences from a large part of the network, which makes this graph hard for the Katz method.
    \item We attribute the poor performance of CMM to the way in which its objective function is designed. The CMM objective function is designed in a way that, in the pursuit of optima, it prefers hyperedges of extremely low cardinality over the rest of the hyperedges. If the candidate hyperedge set has distractor hyperedges of low cardinality, then CMM chooses these hyperedges over genuine high cardinality hyperedges. In real-world networks, more often than not, we observe high hyperedge cardinality (refer Tables \ref{datasets} and \ref{temporal_datasets}).
    
\end{itemize}

\textbf{Ablation Study:}
Our definition of \textit{HRA} has two parts: similarity due to direct connections and due to common neighbors. To analyze the effect of each part, we introduce a weight $\alpha$, and modify the \textit{HRA} equation as follows:
\begin{equation*}
    HRA_{x y} = \alpha HRA_{direct} + (1-\alpha)HRA_{indirect}
\end{equation*}

We vary $\alpha$ over [0,1] to examine the effect of each part on the $AUC$ score (Figure \ref{fig:varying-alpha}). We observe low $AUC$ scores at both extremes, which reveals that both parts are essential in precisely predicting the hyperedges. 

%% file: sections/conclusion.tex
\section{CONCLUSION}
In this work, we have considered the problem of hyperedge prediction. The inherent complexity of the problem makes it difficult to extend existing edge prediction methods for hypergraphs. We proposed \textit{HPRA}, which is the first method that predicts novel hyperedges without any candidate hyperedge set. We accomplish this by efficiently exploiting the hypergraph structure, unlike the existing algorithms, which formulate it as a classification problem. \textit{HPRA} rightly captures the similarity between nodes by extending the \textit{resource allocation} procedure to hypergraphs. Experimental results show that \textit{HPRA} gives unprecedently robust performance compared to state-of-the-art methods. 

\begin{acks}
We are thankful to Mitesh Khapra for his feedback. This work was partially supported by Intel research grant RB/18-19/CSE/002/INTI/BRAV to BR.
\end{acks}

\begin{figure}[ht!]
\centering
\begin{minipage}{0.45\textwidth}
\centering
\includegraphics[width=0.95\linewidth]{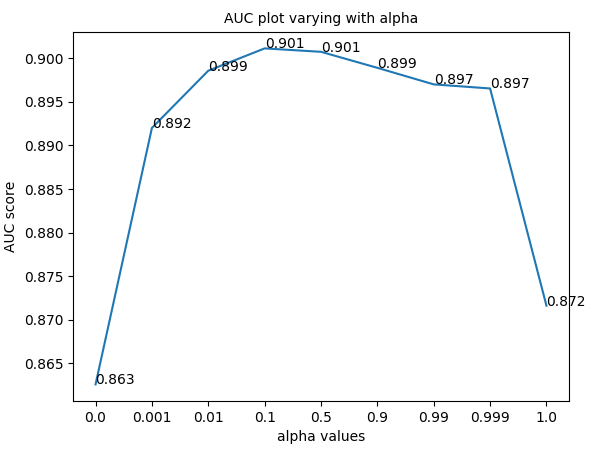}
Citeseer Coreference
\end{minipage}

\begin{minipage}{0.45\textwidth}
\centering
\includegraphics[width=0.95\linewidth]{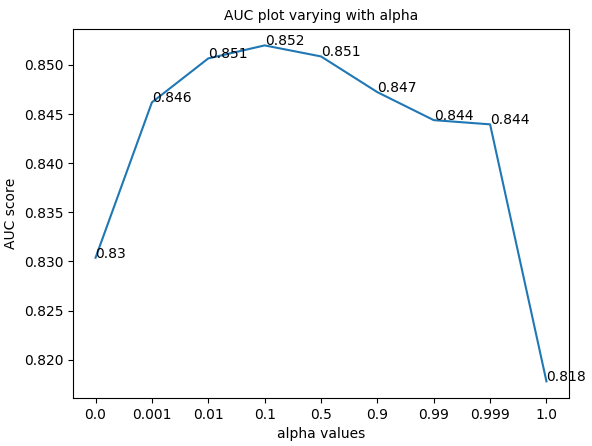}
Cora Coreference
\end{minipage}

\caption{AUC scores vs $\alpha$. We observe lower performance at both extremes implying that both \textit{HRA\textsubscript{direct}} \& \textit{HRA\textsubscript{indirect}} are essential.}

\label{fig:varying-alpha}
\end{figure}